\documentclass[conference]{IEEEtran}
\IEEEoverridecommandlockouts
\usepackage[dvips]{graphicx}
\usepackage[english]{babel}
\usepackage{algorithm}
\usepackage{algorithmic}
\usepackage{amsfonts}
\usepackage{booktabs}
\usepackage[numbers,sort&compress]{natbib}
\usepackage{bm}
\usepackage{enumerate}
\usepackage{amsfonts,amssymb,amsmath}
\usepackage{multirow}
\usepackage{epstopdf}
\usepackage{stfloats}
\usepackage{caption}

\begin{document}

\title{How to Upgrade Wireless Networks: Small Cells or Massive MIMO?\vspace{-0.4cm}}

\author{\IEEEauthorblockN{Xiangxiang Xu$^\dagger$, Xiujun Zhang$^\dagger$, Walid Saad$^*$, Yifei Zhao$^\dagger$ and Shidong Zhou$^\dagger$} \thanks{This work has been partially supported by Science Fund for Creative Research Groups of NSFC (61021001),  Major State Basic Research Development of China (973 program) (2012CB316000), National High Technology Research and Development Program of China (863 program) (2014AA01A703), National Science and Technology Major Project of the Ministry of Science and Technology of China (2013ZX03001024-004), International S\&T Cooperation Program (2012DFG12010), Tsinghua-Qualcomm Joint Research Program, and the U.S. National Science Foundation under Grants CNS-1253731 and CNS-1406947.}
\IEEEauthorblockA{\small $^\dagger$ State Key Laboratory on Microwave and Digital Communications,\\
Tsinghua National Laboratory for Information Science and Technology,\\
Department of Electronic Engineering, Tsinghua University, Beijing 100084, China\\
$^*$ Wireless@VT, Bradley Department of Electrical and Computer Engineering, Blacksburg, VA, USA.\\
Emails: xuxxmail@163.com, zhangxiujun@tsinghua.edu.cn, walids@vt.edu, zhaoyifei@tsinghua.edu.cn, zhousd@mail.tsinghua.edu.cn}\vspace{-1cm}}

\maketitle
\vspace{-80pt}
\begin{abstract}
Radio network deployment and coverage optimization are critical to next-generation wireless networks. In this paper, the problem of optimally deciding on whether to install additional small cells or to upgrade current macrocell base stations (BSs) with massive antenna arrays is studied. This integrated deployment problem is cast as a general integer optimization model by using the facility location framework. The capacity limits of both the radio access link  and the backhaul link are considered. The problem is shown to be an extension of the modular capacitated location problem (MCLP) which is known to be NP-hard. To solve the problem, a novel deployment algorithm that uses Lagrangian relaxation and tabu local search is proposed. The developed tabu search is shown to have a two-level structure and to be able to search the solution space thoroughly. Simulation results show how the proposed, optimal approach to upgrading an existing wireless network infrastructure can make use of a combination of both small cells and BSs with massive antennas. The results also show that the proposed algorithm can find the optimal solution effectively while having a computational time that is up to $30\%$ lower than that of conventional algorithms.
\end{abstract}

\IEEEpeerreviewmaketitle
\vspace{-10pt}
\section{Introduction}
\vspace{-5pt}
 Recently, two paradigms for boosting wireless capacity have emerged: a dense deployment of low-cost, low-power small cell base stations \cite{Quek13} or a massive deployment of antennas at existing base stations (BSs), yielding massive MIMO stations \cite{Marzetta13}. Although these two paradigms have been well studied individually, it is still unclear which one is the most suitable solution for improving capacity and coverage under a given distribution of traffic.

Network planning and deployment has been widely studied in existing literature. Base station deployment was studied in \cite{Fred12} and solved based on force fields. Potential sites for BSs are assumed to be any point within the considered area. In \cite{Peng06}, BSs were deployed to satisfy user demand at minimum cost. However, the interference among different BSs was not accounted for. Location problems for minimizing power consumption were proposed in \cite{Gonzalez11} in which inter-cell interference was neglected so that the transmission rate can be expressed only via power requirement. The study of network deployment without interference was also done in \cite{Guruprasad11} based on a generalized Voronoi partition framework. The works in \cite{Khalek10} and \cite{Amaldi03} studied BS deployment in conventional CDMA systems. Continuous deployment was studied in a homogeneous network in \cite{Khalek10} without accounting for backhaul constraints. In \cite{Amaldi03}, power-based and signal-to-interference ratio (SIR) based uplink transmission were studied and the authors developed a heuristic algorithm using a greedy randomized adaptive search procedure (GRASP) and tabu search (TS) \cite{Glover86}.

While this existing body of work studies interesting deployment problems, several practical wireless challenges such as intercell interference, SIR requirements, and the limitations on the backhaul, were not addressed. More importantly, most of these existing works focused on conventional cellular networks and standard base stations. In contrast, in this work, we focus on whether to deploy small cell base stations or to upgrade to massive MIMO base stations, under backhaul constraints and depending on the network structure. To our best knowledge, this is the first work that addresses this problem.

The main contribution of this paper is to introduce a novel network planning model for cellular downlink transmission in which multiple types of BSs, small cells or massive MIMO, can be deployed, given the network's backhaul constraints. In particular, we model the deployment problem as a facility location problem with inter-cell interference and propose a fast algorithm that jointly integrates techniques from Lagrangian relaxation and tabu search to find a feasible solution. For the proposed algorithm, we show that, since some constraints are relaxed through Lagrangian relaxation, a lower bound of the original problem can be obtained. Then, this lower bound is improved by a subgradient method in an iterated manner. In each iteration, a tabu search algorithm is developed to find feasible solutions started from the solution of the relaxed problem through which an upper bound for the original problem can be computed which is then used to control the update of Lagrangian multipliers. The final solution can be evaluated by studying the gap between the lower and upper bounds obtained. Simulation results assess the various properties of the proposed algorithms and show an insightful example deployment scenario.

The rest of this paper is organized as follows. Section II presents the problem formulation. The proposed solution is discussed in Section III. Section IV gives simulation results and conclusions are drawn in Section V.
\vspace{-5pt}
\section{Problem Formulation}
\vspace{-5pt}
Consider a geographical area $\cal A$ in which a number of small cells must be deployed or a number of macrocell BSs must be upgraded. A set of sites ${\cal F}=\{ 1,\ldots,F\}$ is used to denote the locations of existing macro BSs and potential small cell sites. At site $i \in \cal{F}$, the set of possible BSs is given by ${{\cal K}_i}=\{1,\cdots,K_i \}$. ${{\cal F}_m}=\{ 1,\ldots,F_m\}$ with ${{\cal F}_m} \subseteq {\cal F}$ denotes the set of sites with existing macrocell BSs. For each site $i\in {{\cal F}_m}$, ${{\cal K}_i}$ contains existing macrocell BSs and massive MIMO BSs. For each site $i\in {{\cal F}/{\cal F}_m}$, ${{\cal K}_i}$ contains small cell BSs. Let $c_{ki}$, $C_{ki}$ and $P_{ki}$ be the cost, maximum transmission capacity, and maximum transmit power of facility $k \in {{\cal K}_i}$ at site $i \in {\cal F}$. The backhaul capacity limit is denoted as $C^\textrm{b}_{i}$ for BSs at $i$. We define deployment variables $y_{ki}$ as follows:
\[
\begin{array}{l}
 {y_{ki}} = \left\{ \begin{array}{l}
 1\quad \textrm{if facility } k \textrm{ is deployed at site } i, \\
 0\quad \textrm{otherwise.} \\
 \end{array} \right.
 \end{array}
 \]

Our goal is to find the optimal network deployment strategy which can cover users in area $\cal A$ with minimum cost by either deploying new small cells or upgrading existing macrocell BSs to massive MIMO BSs. The most diffused demand nodes model \cite{Tutschku96} is adopted in which traffic distribution is first discretized  into demand nodes and then demand nodes are treated as mobile users. We denote the set of users by ${\cal S}=\{1,\ldots,S\}$ and we define connection variables $x_{ij}^k$ for $i \in {\cal F}, k \in {\cal K}_i, j \in {\cal S}$ as follows:
 \vspace{-5pt}
\begin{equation*}
\begin{array}{l}
 {x_{ij}^k} = \left\{ \begin{array}{l}
 1\quad \textrm{if user } j \textrm{ is connected to facility } k \textrm{ at site }i,  \\
 0\quad \textrm{otherwise.} \\
 \end{array} \right. \\
 \end{array}
\end{equation*}

 A user $j$ is said to be \emph{served} when its SIR requirement $\gamma_{j}$ is satisfied by a deployed BS whose capacity limit is not violated. We assume that each user can only be served by exactly one BS and, thus, user demand is unsplittable. Formally, the problem can be formulated as the following integer programming.
\vspace{-4pt}
\begin{equation}
\mathop {\min }\limits_{\bm{x},\bm{y}} \;\;\sum\limits_{i=1}^F {\sum\limits_{k=1}^{K_i} {{y_{ki}}{c_{ki}}} },  \label{PObj}
\end{equation}
\vspace{-10pt}
\begin{align}
\textrm{s.t.}\quad &\sum\limits_{k=1}^{K_i} {{y_{ki}}}  \le 1  \qquad i \in {\cal F}, \label{OneOpen} \\
&{x_{ij}^k} \le {y_{ki}} \qquad i \in {\cal F},k \in {{\cal K}_i},j \in {\cal S}, \label{OpenConnected} \\
&\sum\limits_{i=1}^F {\sum\limits_{k=1}^{K_i} {{x_{ij}^k}} }  = 1 \qquad j \in {\cal S}, \label{MustServed}\\
&\sum\limits_{i=1}^F {\sum\limits_{k=1}^{K_i} {{x_{ij}^k}{P_{ki}}h_{ij}^k} } \! \ge\! \gamma_j \sum\limits_{i=1}^F {\sum\limits_{k =1}^{K_i} {({y_{ki}}\! -\! {x_{ij}^k})l_{ki}{{P_{ki}}E_{ki}h_{ij}^k}} },  \label{SIRConstraint}\\
&\sum\limits_{j=1}^S {{r_j}{x_{ij}^k}}  \le {C_{ki}} \qquad i \in {\cal F},k \in {{\cal K}_i}, \label{CapacityLimit}\\
&\sum\limits_{j=1}^S {{r_j}{x_{ij}^k}}  \le {C^\textrm{b}_{i}} \qquad i \in {\cal F}, \label{BackhaulLimit} \\
&\sum\limits_{k = 1}^{{K_i}} {{y_{ki}}}  = 1 \qquad i \in {{\cal F}_m}, \label{MacroMustOpen} \\
&{y_{ki}},{x_{ij}^k} \in \{ 0,1\}.  \label{Integer}
\end{align}
Here, $\bm{x}=[x_{ij}^k](i \in {\cal F}, k\in {\cal K}_i, j \in {\cal S})$ and $\bm{y}=[\bm{y}_m \; \bm{y}_s]=[y_{ki}]( i \in {\cal F}, k \in {\cal K}_i)$ are the optimization parameters. $\bm{y}_m$ and $\bm{y}_s$ are solution vectors denoting the status of macrocell BSs and small cells, respectively. $h_{ij}^k$ is the channel fading from BS $k$ at $i$ to user $j$ and $r_j$ denotes user $j$'s traffic demand. The parameter $0 \le l_{ki} \le 1$ indicates the load condition of facility $k$ at site $i$ which can influence the average interference to other cells. $l_{ki}$ can be defined as $\sum\limits_{j = 1}^S {{r_j}x_{ij}^k}\left/C_{ki}\right.$ (see \cite{Majewski10})
where $\sum\limits_{j=1}^n{r_jx_{ij}^j}$ is the user demand assigned to BS $k$ at location $i$. Here, $l_{ki}$ is the average load of the considered BS and the interference on the right hand side of (\ref{SIRConstraint}) is the averaged inter-cell interference. $E_{ki}\le 0$ is an interference suppression factor. We also note that the capacity limit of each BS is decided not only by the access capacity ${C}_{ki}$ but also by the backhaul transmission constraints $C^\textrm{b}_{i}$. For small cell BSs, the data rate can be low due to a poor backhaul link even when the users experience a good wireless channel.

The objective function in (\ref{PObj}) represents the total installation cost of opened facilities. Constraint (\ref{OneOpen}) implies that at most one facility can be opened at one site and (\ref{OpenConnected}) indicates that mobile users can only be served by opened BSs. (\ref{MustServed}) and (\ref{SIRConstraint}) ensure that every user must be served at a desired SIR level. (\ref{CapacityLimit}) and (\ref{BackhaulLimit}) capture the fact that the total served demand of a BS cannot exceed its access capacity limit and backhaul capacity limit, respectively. (\ref{MacroMustOpen}) implies that macrocell BSs are always open.

We note that, in some instances, constraint (\ref{SIRConstraint}) can be infeasible due to high interference among different BSs. In order to cope with this situation, we reformulate the problem by relaxing constraints (\ref{MustServed}) and (\ref{SIRConstraint}) which require that each user must be served with a SIR higher than a threshold. We relax the constraints such that only users whose quality-of-service (QoS) is satisfied are served and we then aim to maximize the fraction of covered users. To achieve this goal, a term pertaining to the coverage ratio should be added to the objective function. In addition, parameters $l_{ki}$ increase complexity significantly. For reducing complexity, the worst-case scenario is studied in which inter-cell interference always exists with $l_{ki}=1$. So the optimal value in our problem can be regarded as a lower bound of actual performance. Thus, the reformulated problem is:
\begin{equation}
\textrm{(P)}\;\;\;\;\; \mathop {\min }\limits_{\bm{x},\bm{y}} \sum\limits_{i = 1}^F {\sum\limits_{k = 1}^{{K_i}} {{y_{ki}}{c_{ki}}} }  - w\sum\limits_{j = 1}^S {\sum\limits_{i = 1}^F {\sum\limits_{k = 1}^{{K_i}} {r_jx_{ij}^i} } }, \nonumber
\end{equation}
\vspace{-5pt}
\begin{align}
s.t. \quad & (\ref{OneOpen}),(\ref{OpenConnected}),(\ref{CapacityLimit}),(\ref{BackhaulLimit}),(\ref{MacroMustOpen}),(\ref{Integer}),\nonumber\\
&\sum\limits_{i=1}^F {\sum\limits_{k=1}^{K_i} {{x_{ij}^k}} } \le 1 \qquad j \in {\cal S},  \label{ServedOrNot}
\end{align}
\vspace{-5pt}
\begin{equation}\small
\begin{split}
(1-\sum\limits_{i=1}^F {\sum\limits_{k=1}^{K_i}{x_{ij}^k}})M+\sum\limits_{i=1}^F {\sum\limits_{k=1}^{K_i}{x_{ij}^kP_{ij}^k}} \ge \gamma_j \sum\limits_{i=1}^F{\sum\limits_{k=1}^{K_i}{(y_{ki}-x_{ij}^k)P_{ij}^kE_{ki}}}. \label{NewSIR}\\
\end{split}
\end{equation}
where $w>0$ is a biasing factor and $M$ is a number large enough to ensure that constraint (\ref{NewSIR}) is still satisfied when user $j$ is not covered. $P_{ij}^k$ is equal to ${P_{ki}}h_{ij}$ which corresponds to the received power of user $j$ from BS $k$ at $i$. Here, $w$ can be viewed as a profit factor and solving (P) is just the same as maximizing profit through deployment.

Problem (P) can be cast within the framework of the so-called modular capacitated facility location problem which is known to be NP-complete \cite{Correia03}. Therefore, to address this issue, in the next section, we propose a novel, heuristic algorithm based on Lagrangian relaxation and tabu search.
\vspace{-5pt}
\section{Proposed Algorithm}
\vspace{-5pt}
In this section, a Lagrangian algorithm is proposed to obtain a feasible solution to problem (P). The problem (P) is first relaxed by the Lagrangian relaxation and then solved with a greedy algorithm. The optimal value of the relaxed problem is a lower bound of (P) and a subgradient method is exploited to update the Lagrangian multipliers so as to improve this lower bound in an iterated manner.
Note that the solution provided by the
relaxation algorithm can be infeasible to (P). In order to cope with this situation, in each iteration, a two-level iterated tabu search algorithm is developed to obtain and improve feasible solutions based on solutions provided by the relaxation algorithm. Such feasible solutions can then be utilized to guide the update of Lagrangian multipliers. Each step of the approach is described next.
\subsection{Lagrangian Relaxation}
\vspace{-2pt}
By introducing nonnegative Lagrangian multipliers $\lambda_{1j}$,$\lambda_{2j}$,$j\in {\cal S}$ and relaxing (\ref{ServedOrNot}) and (\ref{NewSIR}), we obtain a new optimization problem ($\textrm{P}_{\bm{\lambda}}$) whose optimal value is $V(\bm{\lambda})-\sum\limits_{j=1}^n{(\lambda_{1j}+M\lambda_{2j})}$ where $V(\bm{\lambda})$ is the optimal value of problem $(\textrm{P}_{\bm{\lambda}}')$.
 \begin{equation}
 \begin{split}
(\textrm{P}_{\bm{\lambda}}') \quad &V(\bm{\lambda})=\mathop {\min }\limits_{\bm{y}} \sum\limits_{i = 1}^m {\sum\limits_{k = 1}^{{K_i}} {{V_{ki}}(\bm{\lambda}){y_{ki}}} },  \nonumber\\
&\textrm{s.t.} \quad  (\ref{OneOpen}),(\ref{MacroMustOpen}),(\ref{Integer}). \nonumber
\end{split}
 \end{equation}
 where $V_{ki}(\bm{\lambda})$ is the optimal value of the following assignment problem (${\textrm{P}}_{\bm{\lambda}}^{ki}$) when facility $k$ is open at site $i$.
 \vspace{-5pt}
 \begin{equation}
 \begin{split}
{\textrm{(P}}_{\bm{\lambda}}^{ki}\textrm{)} \quad \mathop {\min }\limits_{\bm{x}} & \sum\limits_{j = 1}^n {\left[ {{\lambda _{1j}}\! - \!{r_j}w \!-\! (P_{ij}^k \!+\! {\gamma _j}P_{ij}^kE_{ki} \!- \! M){\lambda _{2j}}} \right]} x_{ij}^k+  \nonumber \\
&{c_{ki}}\! +\! \sum\limits_{j = 1}^n {{\lambda _{2j}}{\gamma _j}P_{ij}^kE_{ki}}, \nonumber\\
&\textrm{s.t.} \quad (\ref{CapacityLimit}),(\ref{BackhaulLimit}),(\ref{Integer}).\nonumber
\end{split}
\end{equation}
(${\textrm{P}}_{\bm{\lambda}}^{ki}$) is analogous to the so-called knapsack problem and can be solved in a greedy way. Denote the set of users with non-positive coefficients in (${\textrm{P}}_{\bm{\lambda}}^{ki}$) as ${\cal S}_{ki}$. For user $j\notin {\cal S}_{ki}$, let $x_{ij}^k=0$. For those users belonging to ${\cal S}_{ki}$, we distinguish two cases:
 \begin{enumerate}[1)]
   \item \emph{Case 1:} When the total demand of user set ${\cal S}_{ki}$ is less than the capacity limit $C_{ki}$, the optimal solution can be found by just letting $x_{ij}^k=1$ for $j \in {\cal S}_{ki}$.
   \item \emph{Case 2:} When the total demand of set ${\cal S}_{ki}$ violates the capacity limit, the problem is a knapsack problem. Here, we propose a greedy algorithm to solve this problem. The ratio of coefficient of $x_{ij}^k$ in the objective function to user demand $r_j$ is first computed for each user $j\in {\cal S}_{ki}$. Then, users in ${\cal S}_{ki}$ are sorted in a nondecreasing order of calculated ratios. Finally, users are connected to facility $k$ at site $i$ one at a time according to this order until no more users can be added due to the capacity limit.
 \end{enumerate}

 We note that the optimal value of problem ($\textrm{P}_{\bm{\lambda}}$) is a lower bound of the original problem. This lower bound can be improved through iterated update of Lagrangian multipliers. A subgradient method is exploited in our algorithm to update the Lagrangian multipliers.
\subsection{Tabu Search}
\vspace{-2pt}
The tabu search (TS) approach is an extension of conventional hill climbing local search method which can overcome local optima based on a short-term search memory \cite{Glover86}. Here, we use the TS algorithm considering this advantage and the structure of the solution space. The basic procedure of a general TS is as follows. Given an initial feasible solution ${\bm{y}}$, local transformations are made to generate neighbouring solutions denoted by ${\cal N}(\bm{y})$. Then, the best available solution is chosen as the next solution even if it does not improve the objective function. A set of feasible solutions $\{\bm{y}\}$ can be generated and the best one encountered is considered as the final solution after a maximum number of iterations.

In each iteration, the local move that was made is stored and kept for $M$ iterations where $M$ is the length of tabu list. The purpose is to prevent opposite moves which can lead to cycling of local searches. In our problem, three kinds of local move are defined for small cells: 1) \emph{Close move:} Close an opened facility; 2) \emph{Open move:} Open a closed facility. Only those sites without any opened facility are allowed to open a new facility; 3) \emph{Swap move:} Close an opened facility and install a new facility at empty sites.
However, exploring all possible swap moves would be time consuming. In our location problem, it is reasonable to swap between sites which are close to one another considering our actual neighbourhood structure. For an opened small cell $k$ at site $i$, we only consider $N_{swap}$ empty sites with minimal distance to it in the swap move. Macro BSs do not have a close move or an open move due to constraint (\ref{MacroMustOpen}). We define the swap move for macro BSs as follows: 1) \emph{Swap move at the same site:} Change the type of a macrocell BS; 2) \emph{Swap move between different sites:} Change the type of two macro BSs simultaneously.

Although tabu search overcomes local optima, it is still a local search method which allows to explore only a restricted portion of the solution space. Therefore, a so-called diversification mechanism is needed to force the search into unexplored areas. Restart diversification is used in our problem. This approach involves opening $N_{div}$ rarely opened facilities if the best value does not improve for $N_{ni}$ iterations. The search is then restarted from this new solution while the tabu list is simultaneously initialized.

\begin{table}[t]\small
\begin{tabular}
{p{0.95\columnwidth}}
\\
\toprule
\textbf{Algorithm 1} Two-level tabu search algorithm.\\
\midrule
\textbf{Input:} feasible solution ${\bm{y}}$, maximum iteration number $N^{max}_{t_1},N^{max}_{t_2}$, $N_{swap}$, $N_{div}$, tabu list length $M$
\begin{enumerate}[  1:]
\item initialize ${\bm{y}}_0=\bm{y}$, $t_1=0,t_2=0$, best solution $\bm{y}_B=\bm{y}_0$, upper bound $U=V_P(\bm{y}_B)$, empty tabu list
\item \textbf{while($t_1<N^{max}_{t_1}$)}
\end{enumerate}
\begin{enumerate}[  1:\quad]
\setcounter{enumi}{2}
\item compute ${\cal N}({{\bm{y}}_{t_1}})$ through swap moves for macrocell BSs and obtain best solution $\bm{y}_b$ and $V_P(\bm{y}_b)$
\item aspiration criterion: if $V_P(\bm{y}_b)<U$, $\bm{y}_B=\bm{y}_b$, $U=V_P(\bm{y}_b)$, $\bm{y}_n=\bm{y}_b$  \label{Itera1}
\item if $V_P(\bm{y}_b)>U$, find the best nontabu solution and denote as $\bm{y}_n$ \label{Itera2}
\item $t_1=t_1+1$, update tabu list
\item \textbf{while($t_2<N^{max}_{t_2}$)}
\end{enumerate}
\begin{enumerate}[  1:\qquad]
\setcounter{enumi}{7}
\item compute ${\cal N}({{\bm{y}}_{n}})$ through local moves for small cells and obtain best solution $\bm{y}_b$ and $V_P(\bm{y}_b)$
\item repeat procedure \ref{Itera1} and \ref{Itera2}
\item $t_2=t_2+1$ and $\bm{y}_{t_2}=\bm{y}_n$, update tabu list
\item diversification: if best value do not improve for $N_{ni}$ steps, open $N_{div}$ rarely opened facilities, initialize tabu list
\end{enumerate}
\begin{enumerate}[  1:\quad]
\setcounter{enumi}{11}
\item \textbf{end while}
\item $\bm{y}_{t_1}=\bm{y}_n$ and $t_2=0$
\end{enumerate}
\begin{enumerate}[  1:]
\setcounter{enumi}{13}
\item \textbf{end while}
\end{enumerate}
\textbf{Output:} optimized solution $\bm{y}_B$ and optimized value $U$. \\
\bottomrule
\end{tabular}
\end{table}
We propose a two-level TS algorithm considering the interference structure. In our algorithm, the user connection variables $\bm{x}$ and the objective function in (P) are first computed given a BSs' deployment $\bm{y}$. Given $\bm{y}$, each user is first assigned to the opened BS having the strongest receive power. Then, each user's SIR requirements and the capacity limit of each BS are checked. User $j$ may not be served if its SIR does not satisfy threshold $\gamma_j$. For BSs with more demand than capacity, connected users are sorted by demand in a nondecreasing order and disconnected one at a time until capacity limits are satisfied so that a feasible solution $\bm{x}$ can be obtained. This scheme also guarantees that users with more demand have higher priority than users with low demand. Under a given $\bm{y}$, denote the objective value in (P) by $V_{P}(\bm{y})$. The proposed two-level local search process can be stated as follows. Given a BSs' deployment strategy  ${\bm{y}}=\left[\bm{y}_m\;\bm{y}_s\right]$, the neighbourhood ${\cal N}(\bm{y}_m)$ of macro BSs is first evaluated with fixed $\bm{y}_s$ and a swap move is performed for macro BSs. Thus, another feasible solution ${\bm{y}'}=\left[{\bm{y}_m'}\;\bm{y}_s\right]$ is obtained. Then, we search the neighbourhood ${\cal N}(\bm{y}_s)$ of small cells with fixed ${\bm{y}_m'}$ and choose the best non-tabu move after which one local search is finished. The two-level structure is also a diversification scheme which enables one to escape from the neighbourhood ${\cal N}(\bm{y}_s)$ of small cells $\bm{y}_s$ thus leading to a more thorough search of the entire solution space. Here, the algorithm gives the best encountered solution. Since (P) is a minimization programming, the best feasible solution obtained provides an upper bound (UB) for it. The pseudo-code of the two-level tabu search algorithm is given in Algorithm 1.\\
\vspace{-15pt}
\subsection{Proposed heuristic algorithm}
\vspace{-2pt}
Algorithm 2 presents the proposed algorithm which combines Lagrangian relaxation with TS. In algorithm 2, $s$ is the step length used for updating the Lagrangian multipliers, $N_{max}$ is the maximum number of main iterations, ${L}_t$ is the minimum value of the relaxed problem ($\textrm{P}_{\bm{\lambda}}$) in iteration $t$, $\epsilon$ is a gap threshold. $L$ and $U$ denotes the best lower bound and upper bound of problem (P) respectively. $\bm{y}$ and $\bm{x}$ are the final solutions obtained.\\
\begin{table}\small
\begin{tabular}{p{0.95\columnwidth}}
\\
\toprule
\textbf{Algorithm 2} Proposed heuristic algorithm\\
\midrule
\begin{enumerate}[  Step 1:]
\item initialize $t=1$, $U=+\infty$, $L=-\infty$, $s$, $q=0$, $N_1$ and $N_2$.
\item solve problem ($\textrm{P}_{\bm{\lambda}}$) and obtain optimal solution ${\bm{y}}_t$ and optimal value $L_t$. \label{Start}
\item if $L_{t}>L$, $L=L_{t}$ and $q=0$; otherwise $q=q+1$.
\item if $q=N_1$, $s=s/2$ and $q=0$.
\item apply the two-level tabu search from ${\bm{y}}_t$ and update $U$, $\bm{y}$ and $\bm{x}$.
\item if $(U-L)/\left|U\right|< \epsilon$ or complementary slackness is satisfied then stop.
\item update Lagrangian multipliers and $t=t+1$.
\item if $t/N_2$ is integer, initialize $s$.
\item if $t \le N_{max}$ then return to step \ref{Start}.
\end{enumerate}
\textbf{Output:}  $\bm{y},\bm{x}$,$L$ and $U$.
\\
\bottomrule
\end{tabular}\\
\end{table}
\vspace{-10pt}
\section{Simulation Results And Analysis}
\vspace{-5pt}
In our simulations, mobile users are distributed uniformly in a $2$~km $\times$ $2$~km square area. The Hata propagation model \cite{Hatay80} is used for the wireless channel.%
\begin{table}[t]
\setlength{\abovecaptionskip}{0pt}
\setlength{\belowcaptionskip}{0pt}
\centering
\caption{\small Simulation setup}
\begin{tabular}{ccc}
\toprule
 Symbol &Meaning &Value   \\
\midrule
$w$ &Biasing factor &0.2  \\
$\gamma_j$  &SIR requirement of users &$8$~dB \\
$P_m$  &Transmit power of a macrocell BS &$46$~dBm  \\
$P_p$  &Transmit power of a small cell BS &$30$~dBm  \\
$c_m$ &Cost of an existing macro BS &0  \\
$c_p$  &Normalized cost of a small cell BS &1  \\
$c_{massive}$  &Normalized cost of a massive MIMO BS &30 \\
$E_{massive}$  &Interference suppressing factor  &$-20$~dB  \\
 \bottomrule
 \end{tabular}
 \label{SimulationSetup}
\end{table}

In Table \ref{SimulationSetup}, we show the value of the optimization parameters. Furthermore, each BS transmits using its the maximum power. The radio access capacity of conventional macrocell  BSs and small cell BSs are both set to 100 Mbps while a massive MIMO BS is assumed to have a 5 Gbps capacity limit. The backhaul capacity of small cells is assumed to be uniformly distributed between 50 Mbps and 150 Mbps while macrocell BSs have infinite backhaul capacity due to the availability of fiber connections. The target transmission rate for mobile users follows a uniform distribution from 100 Kbps to 8 Mbps which can vary with different types of used applications. Existing macro BSs are located at positions with coordinates $(500,500),(1500,500),(500,1500)$ and $(1500,1500)$. 120 available sites for small cell BSs are considered in our simulation and are distributed uniformly in the considered area. The performance of the proposed algorithm is compared with the one in \cite{Amaldi03} which is based on GRASP and single-level tabu search. Since this algorithm can only find a feasible solution, we integrate it with our Lagrangian relaxation so as to get a lower bound for problem (P). All the simulations were run on a PC with Intel  Core i5-3230M, a 2.6 GHz internal clock and 4 Gb of RAM memory. All statistical results are averaged over all
possible users' locations using a large number of independent simulation runs.

\begin{figure}[!t]
\setlength{\abovecaptionskip}{3pt}
\setlength{\belowcaptionskip}{0pt}
\centering
\scalebox{0.40}{\includegraphics*[4,3][580,546]{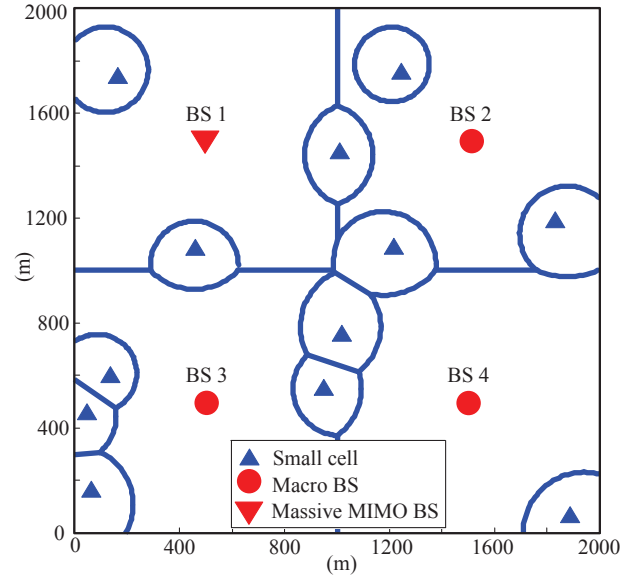}}
\caption{Deployment instance when the number of users is 700.}
\label{DeploymentInstance}
\end{figure}
Fig. \ref{DeploymentInstance} shows a deployment instance of our algorithm when the number of users is 700. In this figure, the locations of deployed small cells and their coverage area are illustrated. Here, 12 small cells are deployed and only one macrocell BS, BS 1, is upgraded to a massive MIMO BS so as to cover the cellular traffic. The upgrade of BS 1 is due to the fact that it has a larger user density within its coverage area. In Fig. \ref{DeploymentInstance}, we can see that small cells are mostly deployed at the cell edge which corroborates the fact that there is a need for improving coverage at the cell edge of macrocell BSs. More importantly, Fig. \ref{DeploymentInstance} shows how massive MIMO and small cells can complement one another. In particular, we can see that most small cells are deployed at the cell edges of macro BSs which were not upgraded to a massive MIMO. Indeed, Fig. \ref{DeploymentInstance} shows that, for the BS that was upgraded to massive MIMO only 3 small cells are needed at its cell edges to optimize coverage. Clearly, upgrading current networks will benefit from a combined deployment of small cells and massive MIMO.

\begin{figure}[!t]
\setlength{\abovecaptionskip}{3pt}
\setlength{\belowcaptionskip}{0pt}
\centering
\scalebox{0.41}{\includegraphics*[7,6][548,380]{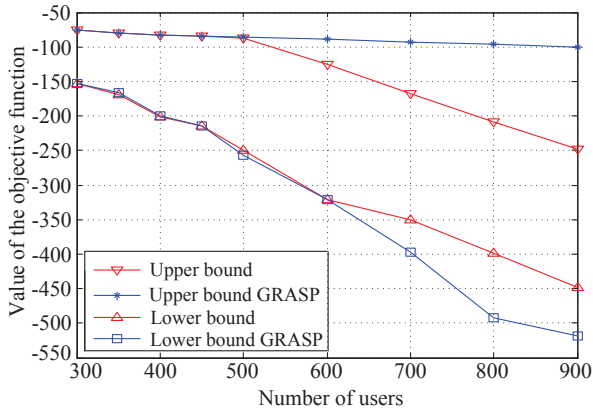}}
\caption{Upper bound and lower bound of problem (P).}
\label{UBLB}
\end{figure}

Fig. \ref{UBLB} shows the upper bounds and lower bounds resulting from our proposed algorithm and the GRASP algorithm in \cite{Amaldi03} under different user traffic density. Fig. \ref{UBLB} shows that, when the number of users is small, the two algorithms obtain the same feasible solution and approximate lower bound. However, our proposed algorithm starts to provide a better deployment solution as the users' density increases. This demonstrates that our two-level structure promotes a more thorough search within the whole solution space and is designed to find better solutions during search process. Fig. \ref{UBLB} also shows that the proposed algorithm provides a better lower bound.

\begin{figure}[!t]
\setlength{\abovecaptionskip}{3pt}
\setlength{\belowcaptionskip}{0pt}
\centering
\scalebox{0.42}{\includegraphics*[5,5][545,365]{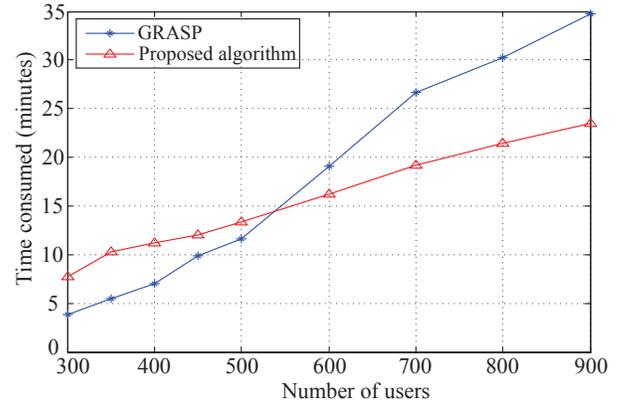}}
\caption{Time consumption of the proposed algorithm.}
\label{TimeConsumption}
\end{figure}

 In Fig. \ref{TimeConsumption}, we compare the time consumption of the proposed algorithm and the reference algorithm. Fig. \ref{TimeConsumption} shows that the proposed algorithm consumes more time when the nework size is small. However, as the number of users increases, the time consumed by GRASP becomes higher than that of the proposed algorithm. The proposed algorithm exhibits a higher efficiency especially for large scale programming. When the number of users is 900, (P) has 115328 binary variables and our algorithm consumes about 24.2 minutes which reduces the overall time consumption of up to 30\% compared with GRASP.

\vspace{-5pt}
\section{Conclusions}
\vspace{-5pt}
In this paper, we have proposed a novel approach to upgrade wireless networks via a potential joint deployment of small cells and massive MIMO upgrades. A general integer optimization model based on the facility location framework has been proposed for optimizing cellular deployment with multiple types of BSs. The proposed model jointly considers the users signal-to-interference ratio, the costs of deployment, as well as backhaul and capacity constraints. To solve this problem, we have proposed a heuristic algorithm by combining a Lagrangian relaxation and a two-level tabu search in an iterated manner. Simulation results have shown that our algorithm can outperform other state-of-the-art approaches. The results have also shown that upgrading existing cellular network infrastructure will require a joint co-deployment of both small cell and massive MIMO technologies.

\small
\def\baselinestretch{0.78}
\bibliographystyle{IEEEtran}
\bibliography{IEEEabrv,refs_paper}
\end{document}